\documentclass[preprint,aps,floats,amssymb,superscriptaddress]{revtex4}

\usepackage{epsfig}

\newcommand{\fpi}{f_\pi}
\newcommand{\mpi}{m_\pi}
\newcommand{\gev}{\, {\rm GeV}}
\newcommand{\mev}{\, {\rm MeV}}
\newcommand{\fm}{\, {\rm fm}}
\newcommand{\non}{\nonumber}
\newcommand{\tr}{{\rm tr\,}}
\newcommand{\Lg}{{\cal L}}

\newcommand{\str}{{\rm \, str}}

\begin{document}

\preprint{
\vbox{
\hbox{\today}
\hbox{ADP-02-73/T512, LTH-537}
}}

\title{Chiral Analysis of Quenched Baryon Masses}

\author{R. D. Young}
\author{D. B. Leinweber}
\author{A. W. Thomas}
\affiliation{    Special Research Centre for the
                 Subatomic Structure of Matter,
                 and Department of Physics and Mathematical Physics,
                 University of Adelaide, Adelaide SA 5005,
                 Australia}

\author{S. V. Wright}
\affiliation{    Special Research Centre for the
                 Subatomic Structure of Matter,
                 and Department of Physics and Mathematical Physics,
                 University of Adelaide, Adelaide SA 5005,
                 Australia}
\affiliation{    Division of Theoretical Physics,
                 Department of Mathematical Sciences, 
                 University of Liverpool,
                 Liverpool L69 3BX, U.K.}

\begin{abstract}
We extend to quenched QCD an earlier investigation of the chiral
structure of the masses of the nucleon and the delta in lattice
simulations of full QCD. Even after including the meson-loop
self-energies which give rise to the leading and next-to-leading
non-analytic behaviour (and hence the most rapid variation in the
region of light quark mass), we find surprisingly little curvature in
the quenched case.  Replacing these meson-loop self-energies by the
corresponding terms in full QCD yields a remarkable level of agreement
with the results of the full QCD simulations. This comparison leads to
a very good understanding of the origins of the mass splitting between
these baryons.

\end{abstract}

\maketitle


\begin{section}{Introduction}

The quenched approximation is a widely used tool for studying
non-perturbative QCD within numerical simulations of lattice gauge
theory. With an appropriate choice of the lattice scale and at
moderate to heavy quark masses, this approximation has been shown to
give only small, systematic deviations from the results of full QCD
with dynamical fermions. Although no formal connection has been
established between full and quenched QCD, the similarity of the
results has led to the belief that the effects of quenching are small
and hence that quenched QCD provides a reasonable approximation to the
full theory \cite{Aoki:1999yr}.

Improved lattice actions, together with advances in high performance
computing, have been responsible for significant improvements in the
calculation of baryon masses at moderate to light quark masses within
the quenched approximation
\cite{Bowler:1999ae,Kanaya:1998sd,Bernard:2001av,Zanotti:2001yb}.
Simulations with dynamical fermions have proven to be more difficult,
but results have been reported with pion masses as low as $320\mev$
\cite{Bernard:2001av,Aoki:1999ff}.

The fact that one is restricted to quark masses much larger than the
physical values means that, in addition to all the usual
extrapolations (e.g., to the infinite volume and continuum limits), if
one wants to compare with empirical hadron observables, one must also
have a reliable method of extrapolation to the chiral limit. Any such
extrapolation must incorporate the appropriate chiral corrections,
arising from Goldstone boson loops, which give rise to rapid,
non-linear variations as the chiral limit is approached. The
importance of incorporating such behaviour has been successfully
demonstrated for a number of hadronic observables, including masses
\cite{Leinweber:1999ig,Leinweber:2001ac}, the sigma commutator
\cite{Leinweber:2000sa}, magnetic moments
\cite{Leinweber:1998ej,Hackett-Jones:2000qk,Leinweber:1999nf,Hemmert:2002uh},
charge radii \cite{Hackett-Jones:2000js} and parton distribution
functions \cite{Detmold:2001jb,Detmold:2001dv}.

The impressive results found using these methods have led us to the
present investigation of the problem of the chiral extrapolation of
baryon masses in quenched QCD. The chiral properties within the
quenched approximation are known to differ from those of full QCD in a
number of very interesting ways
\cite{Sharpe:1990me,Sharpe:1992ft,Bernard:1992mk,Labrenz:1996jy,Leinweber:2001jc,Savage:2001dy}.
For example, not only are the effective couplings at the pion-baryon
vertices significantly altered in quenched QCD (QQCD) but, because the
$\eta'$ behaves as a Goldstone boson in QQCD, one must also consider
$\eta'$ loops.

Here we first review previous work \cite{Leinweber:1999ig} which
reported a successful method for extrapolating baryon masses as
calculated in full QCD lattice simulations. The modified chiral
structure of quenched baryon masses \cite{Labrenz:1996jy} is presented
next. We show how to construct the various meson loop induced
self-energies in order to preserve the leading-non-analytic (LNA) and
next-to-leading non-analytic (NLNA) structure appropriate to QQCD,
while incorporating the established behaviour at heavier quark masses.
This is followed by a detailed application to the extrapolation of the
quenched $N$ and $\Delta$ masses to the chiral limit. Finally, we use
the observed similarity of the structure of baryons stripped of their
Goldstone boson clouds, in full and quenched QCD, to explore whether
one can make a connection between the masses calculated in QQCD and
those obtained in a dynamical simulation. The remarkable agreement
obtained suggests a number of further tests and also leads us, with
considerable confidence, to an interpretation of the origin of the
$N$--$\Delta$ mass splitting.

\end{section}


\begin{section}{QCD Extrapolation}
In general, the coefficients of the LNA and NLNA terms in a chiral
expansion of baryon masses are very large. For instance, the LNA term
for the nucleon mass is $\delta m_N^{(\rm LNA)} = -5.6\, m_\pi^3$
(with $m_\pi$ and $\delta m_N^{(\rm LNA)}$ in GeV). With $m_\pi =
0.5\gev$, quite a low mass for current simulations, this yields
$\delta m_N^{(\rm LNA)} =0.7\gev$ --- a huge
contribution. Furthermore, in this region hadron masses in both full
and quenched lattice QCD are found to be essentially linear in
$m_\pi^2$ or equivalently quark mass, whereas $\delta m_N^{(\rm LNA)}$
is highly non-linear. The challenge is therefore to ensure the
appropriate LNA and NLNA behaviour, {\it with the correct
coefficients}, as $m_\pi \rightarrow 0$, while making the transition
to linear behaviour as $\mpi$ increases, sufficiently rapidly to
describe the actual lattice data.

A reliable method for achieving all this was proposed by
Leinweber~{\it et~al.}~\cite{Leinweber:1999ig}.
They fit the full (unquenched) lattice data with the form:
\begin{equation}
M_B = \alpha_B + \beta_B \mpi^2 + \Sigma_B (\mpi, \Lambda) ,
\label{eq:FullFit}
\end{equation}
where $\Sigma_B$ is the total contribution from those pion loops  
which give rise to the LNA and NLNA terms in the self--energy of the 
baryon. For the $N$ these correspond to the 
processes $N\to N\pi\to N$ and $N\to \Delta\pi\to N$, while for the 
$\Delta$ we need $\Delta \to \Delta \pi\to \Delta$ and 
$\Delta \to N \pi \to \Delta$. Explicitly,
\begin{eqnarray}
\Sigma_N      & = & \sigma_{NN}^\pi + \sigma_{N\Delta}^\pi ,            \non   \\
\Sigma_\Delta & = & \sigma_{\Delta\Delta}^\pi + \sigma_{\Delta N}^\pi .
\end{eqnarray}
In the heavy 
baryon limit, these four contributions ($B\to B'\pi\to B$) can be 
summarised as: 
\begin{equation}
\sigma_{BB'}^\pi =  - \frac{3}{16\pi^2 \fpi^2} G_{BB'} 
\int_0^\infty dk \frac{k^4 u^2(k)}
{\omega(k) ( \omega_{BB'} + \omega(k) )},
\label{eq:fullSEpi}
\end{equation}
where $\omega(k)=\sqrt{k^2+\mpi^2}$ is the intermediate pion energy
and $\omega_{BB'} = (M_{B'} - M_B)$ is the physical baryon mass
splitting.  The coefficients $G_{BB'}$ are standard SU(6) couplings
and are summarised in Section~\ref{sec:se}. The ultraviolet regulator,
$u(k)$, has a very natural physical interpretation as the Fourier
transform of the source of the pion field. The LNA and NLNA structure
of these diagrams is associated with the infrared behaviour of the
corresponding integrals and hence is independent of the choice of
regularisation scheme. The use of a form factor effectively suppresses
the self-energies like $\Lambda^2/\mpi^2$ for $m_\pi \gg \Lambda$, the
characteristic mass scale of the cutoff. A common choice of regulator,
which we use throughout this work, is the dipole form, $u(k) =
\Lambda^4 / (\Lambda^2 + k^2)^2$.

The linear term of Eq.~(\ref{eq:FullFit}), which dominates for $m_\pi
\gg \Lambda$, models the quark mass dependence of the pion-cloud
source --- the baryon without its pion dressing.  This term also
serves to account for loop diagrams involving heavier mesons, which
have much slower variation with quark mass. Given the current state of
the art in lattice simulations, data in the low to intermediate mass
range is unable to reliably constrain the parameter $\Lambda$. There
is considerable phenomenological support for choosing a dipole mass
parameter somewhat smaller than found for the axial form factor of the
nucleon, which is $1.03\pm0.04\gev$
\cite{Thomas:2001kw,Guichon:1983zk,Thomas:1989tv}. However, it is
important to understand that the anticipated development of
supercomputing resources and techniques are such that $\Lambda$ may be
constrained by full QCD simulation data within the next five years.

Fitting lattice results to Eq.~(\ref{eq:FullFit}) is
straightforward. Upon calculating the described self-energies for a
given choice of $\Lambda$, the fitting procedure amounts to a simple
linear fit in $\alpha_B$ and $\beta_B$.

\end{section}


\begin{section}{Quenched Chiral Perturbation Theory}
Standard chiral perturbation theory ($\chi$PT) is a low energy
effective field theory built upon the symmetries of QCD
\cite{Gasser:1984yg,Bernard:1995dp}. It amounts to an expansion of
Green's functions in powers of momenta and quark mass about the chiral
limit ($m_q=0$). In the case of baryon masses, $\chi$PT tells us the
leading behaviour of the quark mass expansion.  Because $\chi$PT is an
effective field theory, the renormalisation procedure must be
performed order by order in perturbation theory. At higher and higher
order, more and more unknown parameters are introduced. These
parameters only play a role in analytic terms of the expansion. The
coefficients of the leading non-analytic terms are constrained by
chiral symmetry \cite{Gasser:1988rb} --- they are independent of
regularisation and the order of the chiral expansion. In connecting
the results of lattice QCD to the physical world it is essential that
one incorporates the correct non-analytic structure of the low energy
theory.

Quenched $\chi$PT (Q$\chi$PT) provides the analogous low energy
effective theory for QQCD
\cite{Sharpe:1992ft,Bernard:1992mk,Labrenz:1996jy}.  Sea quark loops
are removed from QCD by including a set of degenerate, commuting
(bosonic) quark fields.  These bosonic fields have the effect of
exactly cancelling the fermion determinant in the functional
integration over the quark fields.  This gives a Lagrangian field
theory which is equivalent to the quenched approximation simulated on
the lattice.  The low energy effective theory is then constructed on
the basis of the symmetries of this Lagrangian.

The leading chiral expansion of baryon masses in the quenched
approximation has been calculated by Labrenz and Sharpe
\cite{Labrenz:1996jy}. For the reasons already mentioned in the
Introduction, it differs from the corresponding expansion in full QCD.
In particular, the chiral expansion coefficients take different values
and new, non-analytic behaviour is also introduced. The explicit form
can be expressed as
\begin{table}
\begin{center}
\begin{ruledtabular}
\begin{tabular}{llcc}
$B$      & $c_i$    & QCD                       & QQCD                                                \\
\hline
         & $c_1$    & $0$                       & $-\frac{1}{2}(3F-D)^2 m_0^2 \nu$                    \\
$N$      & $c_3$    & $-\frac{3}{2}(F+D)^2 \nu$ & $\left\{ \frac{4}{3}(D^2-3DF) - 2(3F-D)\gamma 
                                              \right\} \nu $                                          \\
         & $c_{4L}$ & $\frac{3}{\pi}(F+D)^2\frac{32}{25}\frac{3}{8\Delta M} \nu$ 
                                                & $\frac{{\cal C}^2}{2\pi \Delta M} \nu$              \\
\hline
         & $c_1$    & $0$                       & $-\frac{5}{18}{\cal H}^2 m_0^2 \nu$                 \\
$\Delta$ & $c_3$    & $-\frac{3}{2}(F+D)^2 \nu$ & $\left\{ -\frac{10}{27}{\cal H}^2 - 
                                                   \frac{10}{9}{\cal H}\gamma' \right\} \nu $         \\
         & $c_{4L}$ & $-\frac{3}{\pi}(F+D)^2\frac{8}{25}\frac{3}{8 \Delta M} \nu$ 
                                                & $\frac{{\cal C}^2}{2\pi \Delta M}\frac{4}{25} \nu$  \\
\end{tabular}
\end{ruledtabular}
\caption{Coefficients of the lowest order non-analytic terms in 
the chiral expansions of the $N$ and $\Delta$ masses, with values from 
both full and quenched QCD listed for comparison ($\nu^{-1}=16\pi\fpi^2$, 
$\Delta M = M_\Delta - M_N$). 
\label{tab:chicoeffs}}
\end{center}
\end{table}
\begin{equation}
M_B = M_B^{(0)} + c_1^B \mpi + c_2^B \mpi^2  + c_3^B \mpi^3 
        + c_4^B \mpi^4 + c_{4L}^B \mpi^4 \log \mpi + \ldots
\label{eq:chiexp}
\end{equation}
with the coefficients of the terms which are non-analytic in the quark
mass listed in Table~\ref{tab:chicoeffs}. We note that in
Ref.~\cite{Labrenz:1996jy} the $N$ and $\Delta$ were treated as
degenerate states in the chiral limit. Experience in other situations
suggests that it is more accurate to retain a finite mass difference,
in which case off--diagonal terms such as $N \rightarrow \Delta \pi
\rightarrow N$ lead to the non-analytic behaviour of the form $\mpi^4
\log\mpi$.

The contribution linear in $\mpi$ is unique to the quenched
approximation.  The quenched theory therefore exhibits a more singular
behaviour in the chiral limit. The origin of this behaviour is the
Goldstone nature of the $\eta'$ in QQCD and specifically the process
shown in Fig.~\ref{fig:etaQF}(b).  We note also that the coefficients
of the chiral expansion involve new couplings, $\gamma$ and $\gamma'$,
which are related to the {}flavour-singlet, hairpin--baryon couplings
for $N$ and $\Delta$ respectively, illustrated in
{}Fig.~\ref{fig:etaQF}(a). In the formalism of
Ref.~\cite{Labrenz:1996jy} these are related to the couplings of full
QCD via the relations,
\begin{equation}
\gamma=D-F\, , \hspace{5mm} \gamma'=0\, ,
\label{eq:gamma}
\end{equation}
as described in Appendix~\ref{app:singlet}. There is some uncertainty
over the flavour singlet couplings, especially in connection with OZI
violation associated with the U(1) axial anomaly
\cite{Bass:1999is}. While this may modify our calculated curves at
extremely light quark mass, it would have no significant effect on the
fit to lattice data at large quark mass nor on the comparison of
current quenched and full QCD data.
\begin{figure}[!t]
\begin{center}
{\epsfig{file=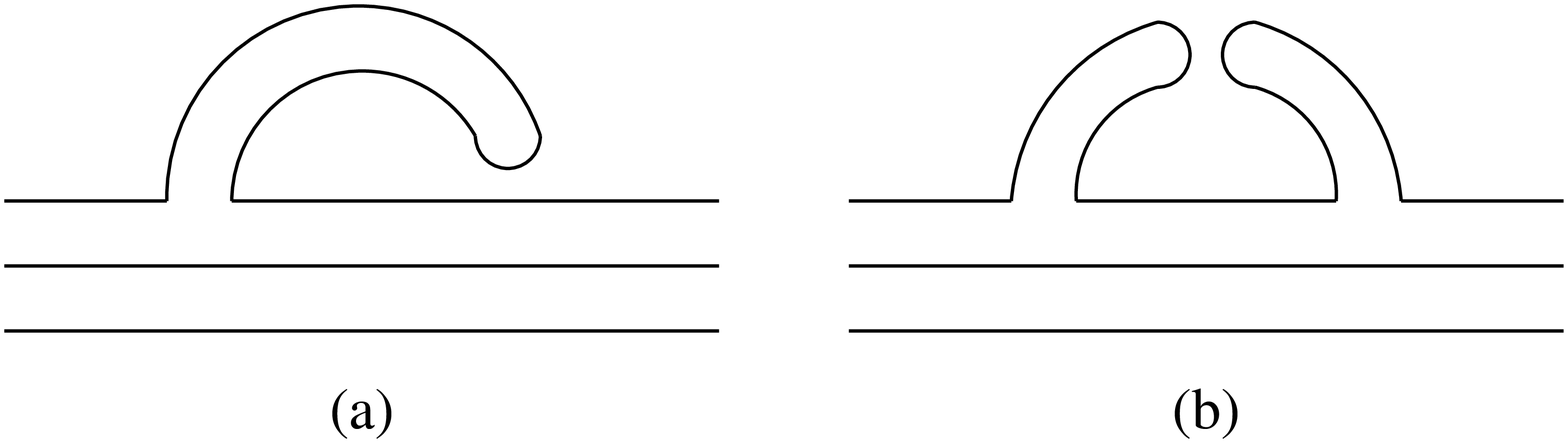, width=11cm, angle=0}}
\caption{Quark flow diagrams for the chiral $\eta'$ loop contributions 
appearing in QQCD: (a) single hairpin, (b) double hairpin.
\label{fig:etaQF}}
\end{center}
\end{figure}

\end{section}


\begin{section}{Quenched Self Energies}
\label{sec:se}
Our aim is to apply a similar procedure for the chiral extrapolation
of quenched QCD data to that which has proven successful for the
physical theory. That is, we wish to generalise Eq.~(\ref{eq:FullFit})
to replace the LNA and NLNA self-energy terms arising in full QCD by
their quenched analogues. The pion loop contributions have the same
kinematic structure as those in full QCD.  A simple redefinition of
the couplings, $G_{BB'}$, in the expressions for the self-energies
ensures that the correct LNA and NLNA of Q$\chi$PT is
maintained. Thus, the analytic expressions for the pion cloud
corrections to the masses of the $N$ and $\Delta$ are of the same form
as the full QCD integrals (c.f.~Eq.~(\ref{eq:fullSEpi})):
\begin{equation}
\tilde{\sigma}^\pi_{BB'}= 
-\frac{3}{16\pi^2 \fpi^2} \tilde{G}_{BB'} \int_0^\infty dk \frac{k^4 u^2(k)}
{\omega(k) ( \omega_{BB'} + \omega(k) )} .
\label{eq:QQCD}
\end{equation}
where the quenched couplings, $\tilde{G}_{BB'}$, are listed in
Table~\ref{tab:Gcoeffs}, together with their physical counterparts.
\begin{table}
\begin{center}
\begin{ruledtabular}
\begin{tabular}{lcccc}
& $G_{NN}$	&  $G_{N\Delta}$		&  $G_{\Delta\Delta}$	&  $G_{\Delta N}$         \\
\hline
QCD	  &
$(F+D)^2$	&   $\frac{32}{25}(F+D)^2$	&  $(F+D)^2$		&  $\frac{8}{25}(F+D)^2$  \\
QQCD	  &
$\frac{8}{9}(3DF - D^2)$	&  $\frac{16}{9}D^2$	&
$\frac{20}{9}D^2$		& $-\frac{8}{9}D^2$ 		\\
\end{tabular}
\end{ruledtabular}
\caption{ Chiral couplings appearing in the self-energy integrals,
Eq.~(\protect\ref{eq:fullSEpi}) for full QCD and
Eq.~(\protect\ref{eq:QQCD}) for QQCD.  In numerical calculations we
have used the couplings arising from $SU(6)$ relations
\protect\cite{Butler:1993pn}, ${\cal C}=-2D$ and ${\cal H}=-3D$, with
the tree-level parameters, $F=0.50$ and $D=0.76$.
\label{tab:Gcoeffs}
}
\end{center}
\end{table}

Within the quenched approximation $\eta'$ loops also contribute to the
low energy effective theory, whereas they are usually neglected in the
physical case. This is because a re-summation of internal loop
diagrams (coming from the fermion determinant) means that the $\eta'$
remains massive in the chiral limit of full QCD. On the other hand,
the absence of these virtual loops in the quenched approximation
causes the flavour singlet $\eta'$ to behave as a Goldstone boson
\cite{Bernard:1992mk,Sharpe:1992ft}. As a consequence of this feature
of the quenched theory, there are two new types of loop contributions
to be considered. A schematic view of these processes is shown in
Fig.~\ref{fig:etaQF}.

The first of these two contributions, shown in
Fig.~\ref{fig:etaQF}(a), arises from a single ``hairpin''
interaction. As discussed above, it is responsible for the term
proportional to $\gamma$ ($\gamma'$) in the chiral expansion of the
$N$ ($\Delta$) mass. These couplings are discussed in considerable
detail in Appendix~\ref{app:singlet}. The structure of this diagram is
exactly the same as the pion loop contribution where the internal
baryon is degenerate with the external state. The integral
representing this diagram is then the same as that for
$\tilde{\sigma}^\pi_{BB}$,
\begin{equation}
\tilde{\sigma}_B^{\eta'(1)} = -\frac{3}{16\pi^2 \fpi^2} N_B^{(1)} \int_0^\infty dk \frac{k^4 u^2(k)}
                        {\omega^2(k)} .
\label{eq:SEeta1}
\end{equation}
The factors $N_B^{(1)}$, providing the correct non-analytic behaviour
in the chiral expansion (Eq.~(\ref{eq:chiexp})), are displayed in
Table~\ref{tab:Nb}.

The second of these new $\eta'$ loop diagrams arises from the double
hairpin vertex, pictured in Fig.~\ref{fig:etaQF}(b). This contribution
is particularly interesting because there are two meson propagators
and it is therefore responsible for the non-analytic term linear in
$\mpi$ --- this term being unique to the quenched case. The integral
corresponding to this self energy can be written in a similar way:
\begin{equation}
\tilde{\sigma}_B^{\eta'(2)} = \frac{3}{16\pi^2 \fpi^2} N_B^{(2)} \int_0^\infty dk \frac{k^4 u^2(k)}
                        {\omega^4(k)} .
\label{eq:SEeta2}
\end{equation}
Note the sign change and the higher power of $\omega$ in the
denominator. The coefficients, $N_B^{(2)}$, providing the correct
non-analytic behaviour in Eq.~(\ref{eq:chiexp}) --- in this case the
coefficient of $m_\pi$ --- are given in Table~\ref{tab:Nb}.
\begin{table}
\begin{center}
\begin{ruledtabular}
\begin{tabular}{lcc}
         & $N_B^{(1)}$                      & $N_B^{(2)}$ (GeV$^2$)           \\
\hline
$N$      & $\frac{4}{3}(3F-D)\gamma$       & $\frac{2}{9}(3F-D)^2 m_0^2$     \\
$\Delta$ & $\frac{20}{27}{\cal H}\gamma'$  & $\frac{10}{81}{\cal H}^2 m_0^2$ \\
\end{tabular}
\end{ruledtabular}
\caption{Couplings used in flavour singlet $\eta'$ self-energies. We
take $m_0^2 = 0.42\gev^2$, lying between phenomenological and lattice
estimates \cite{Kuramashi:1994aj,Bardeen:2000cz,DeGrand:2002gm}. The
momentum dependence of the double hairpin vertex, which is believed to
be small, is neglected.
\label{tab:Nb}}
\end{center}
\end{table}

The sum of these four contributions then gives the net meson-loop induced 
self-energies within the quenched approximation,
\begin{equation}
\tilde{\Sigma}_B = \tilde{\sigma}^\pi_{BB} + \tilde{\sigma}^\pi_{BB'} +
               \tilde{\sigma}_B^{\eta'(1)} + \tilde{\sigma}_B^{\eta'(2)}.
\end{equation}

The individual contributions to the $N$ and $\Delta$ masses over a
range of pion mass are plotted in Figs.~\ref{fig:SEconN} and
\ref{fig:SEconD}. These are all evaluated with the dipole regulator
mass parameter $\Lambda=0.8\gev$. The corresponding self-energies from
full QCD are also shown for comparison.  We note that in QQCD the
contributions are typically quite a bit smaller and the double-hairpin
graph, $\tilde{\sigma}_B^{\eta'(2)}$, is repulsive. The differences
are enhanced for the $\Delta$ where $\tilde{\sigma}^\pi_{\Delta N}$ is
also repulsive. We observe that the rapid, non-linear behaviour (which
is effectively much larger in full QCD) is restricted to the region
$\mpi^2\lesssim 0.2\gev^2$, above which the self-energies are quite
smoothly varying functions of the quark mass.
\begin{figure}[t]
\begin{center}
{\epsfig{file=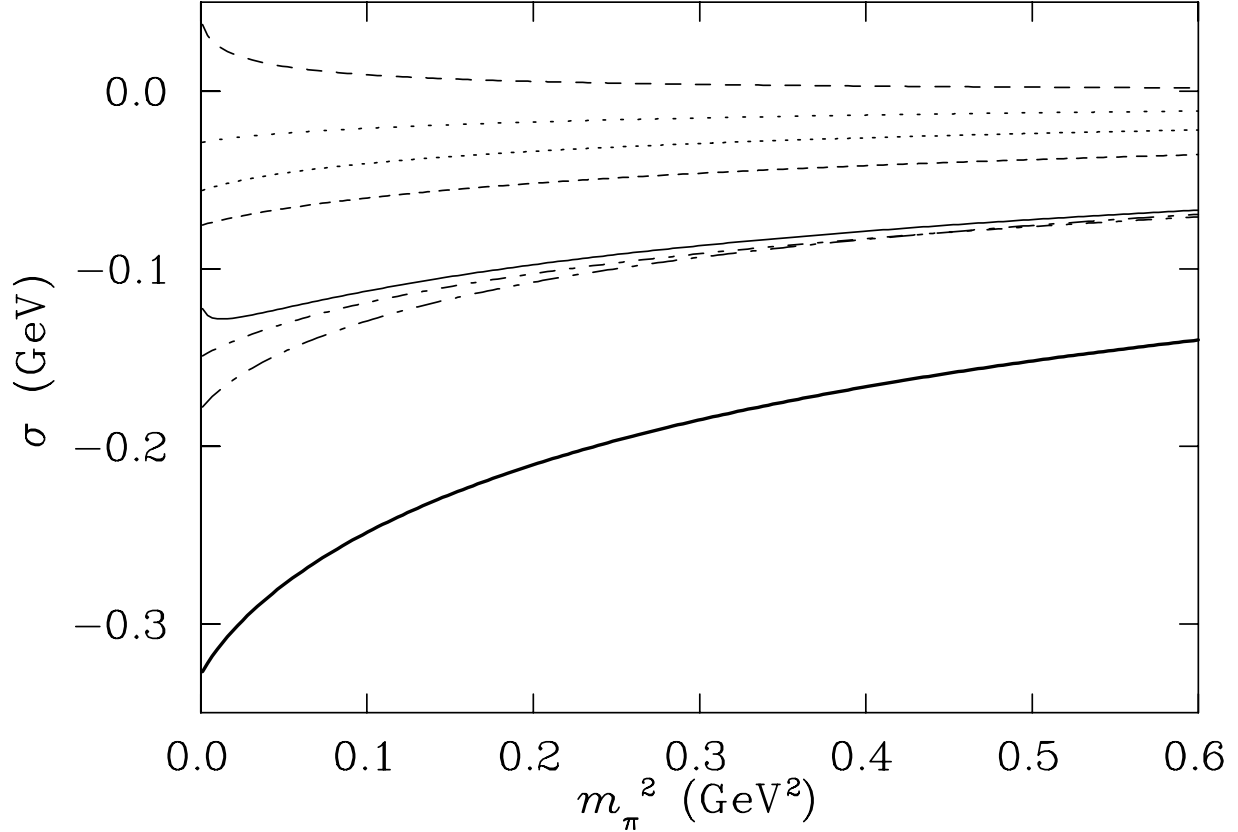, width=11cm, angle=90}}
\caption{Various self-energy contributions to $M_N$ for dipole mass,
$\Lambda=0.8\gev$.  From top down at $\mpi^2=0.1\gev^2$, the curves
correspond to (where a $\sim$ over the symbol denotes a quenched QCD
contribution) 
$\tilde{\sigma}_N^{\eta'(2)}$,
$\tilde{\sigma}_N^{\eta'(1)}$, 
$\tilde{\sigma}^\pi_{NN}$,
$\tilde{\sigma}^\pi_{N \Delta}$, 
total quenched $\tilde\Sigma_N$,
$\sigma^\pi_{N \Delta}$, 
$\sigma^\pi_{NN}$ 
and total physical $\Sigma_N$.
\label{fig:SEconN}}
\end{center}
\end{figure}
\begin{figure}[t]
\begin{center}
{\epsfig{file=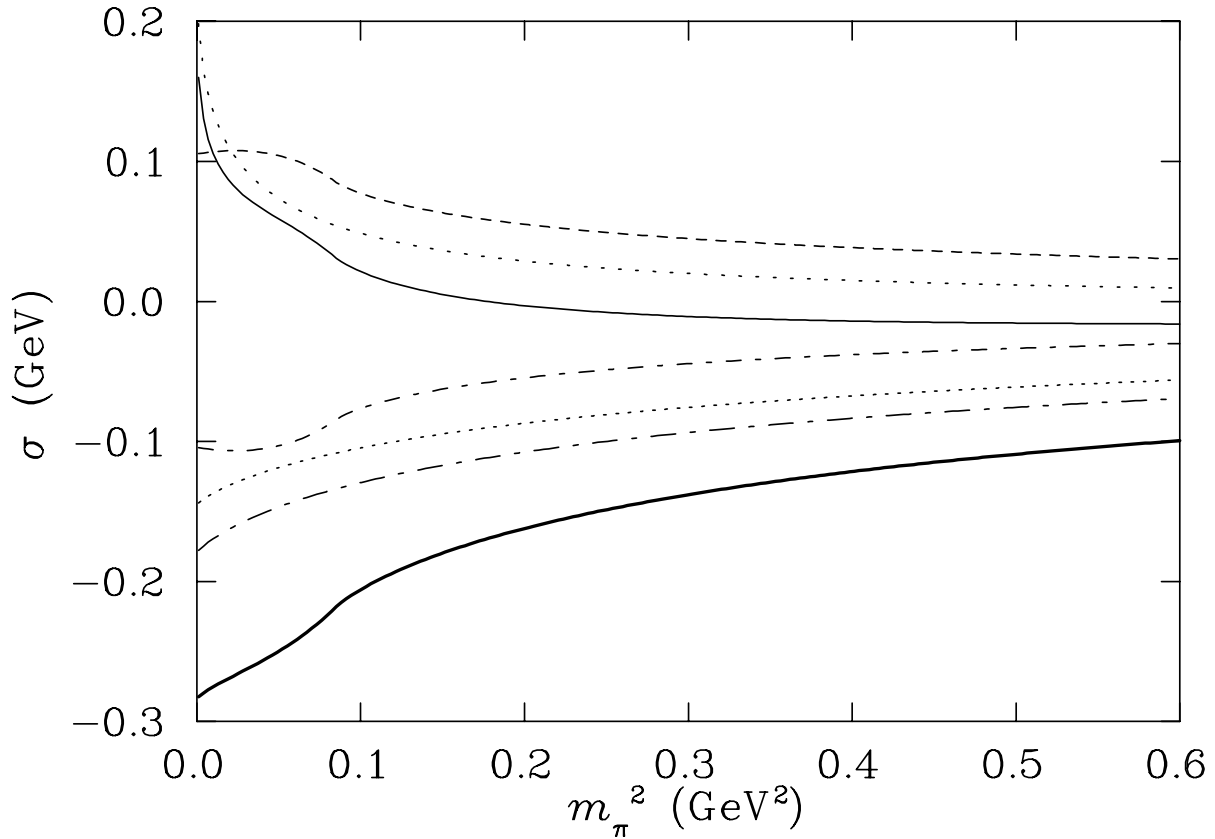, width=11cm, angle=90}}
\caption{Various self-energy contributions to $M_\Delta$ for dipole
mass, $\Lambda=0.8\gev$.  From top down at $\mpi^2=0.1\gev^2$, the
curves correspond to (where a $\sim$ over the symbol denotes a
quenched QCD contribution)
$\tilde{\sigma}^\pi_{\Delta N}$, 
$\tilde{\sigma}_\Delta^{\eta'(2)}$,
total quenched $\tilde\Sigma_\Delta$,
$\sigma^\pi_{\Delta N}$,
$\tilde{\sigma}^\pi_{\Delta\Delta}$, 
$\sigma^\pi_{\Delta\Delta}$
and total physical $\Sigma_\Delta$.
\label{fig:SEconD}}
\end{center}
\end{figure}

\end{section}


\begin{section}{Fitting Procedure}
The lattice data considered in this analysis comes from the recent paper of
Bernard~{\em et~al.} \cite{Bernard:2001av}. These simulations were
performed using an improved Kogut-Susskind quark action, which is
known to have good scaling properties \cite{Bernard:1999xx}.  Unlike
the standard Wilson fermion action, masses determined at finite
lattice spacing are excellent estimates of the continuum limit
results.

We are particularly concerned with the chiral extrapolation of baryon
masses and how their behaviour is affected by the quenched
approximation. In such a study, it is essential that the method of
scale determination is free from chiral contamination. One such method
involves the static-quark potential. As low-lying pseudoscalar mesons
made of light quarks exhibit negligible coupling to hadrons containing
only heavy valence quarks, the low energy effective field theory plays
no role in the determination of the scale for these systems. In fixing
the scale through such a procedure one constrains all simulations,
quenched, 2-flavour, 3-flavour {\em etc.}, to match phenomenological
static-quark forces.  Effectively, the short range ($0.35 \sim
0.5\fm$) interactions are matched across all simulations.

A commonly adopted method involving the static-quark potential is the
Sommer scale \cite{Sommer:1994ce,Edwards:1998xf}. This procedure
defines the force, $F(r)$, between heavy quarks at a particular length
scale, namely $r_0\simeq0.5\, {\rm fm}$. Choosing a narrow window to
study the potential avoids complications arising in dynamical
simulations where screening and ultimately string breaking is
encountered at large separations. The lattice data analysed in this
report uses a variant of this definition, choosing to define the force
at $r_1=0.35\, {\rm fm}$ via $r_1^2 F(r_1)=1.00$
\cite{Bernard:2001av}.

As we remarked earlier, the non-analytic chiral behaviour is governed
by the infrared regions of the self-energy integrals. The fact that
the lattice calculations are performed on a finite volume grid means
that the self-energy integrals implicit in current lattice simulations
do not include the exact chiral behaviour. It is important to take
this into account in the fitting procedure and we therefore follow
Ref.~\cite{Leinweber:2001ac} in replacing the continuum self-energy
integrals used in the fitting process by a discrete sum over the meson
momenta available on the lattice:
\begin{equation}
4\pi\int_0^\infty k^2 dk = \int d^3k \approx \frac{1}{V} \left(\frac{2\pi}{a}\right)^3 \sum _{k_{x},k_{y},k_{z}}\, .
\label{eq:discrete}
\end{equation}
The self-energy integrals calculated in this way are what should be
directly compared with the lattice data, and we illustrate these by
open squares in subsequent figures. Upon obtaining the optimal fit
parameters, one can evaluate the integrals exactly and therefore
obtain the infinite-volume, continuum limit. The latter is the result
which should be compared with experiment at the physical pion mass.

We now proceed to fit quenched lattice data with the form
\begin{equation}
\tilde{M}_B = \tilde{\alpha}_B + \tilde{\beta}_B \mpi^2 + \tilde{\Sigma}_B (\mpi, \Lambda) ,
\label{eq:QnchFit}
\end{equation}
(by analogy with the form used in full QCD, Eq.~(\ref{eq:FullFit})),
with the self-energies evaluated, as we have just outlined, using the
momentum grid corresponding to the specific lattice simulation.  The
linear terms in Eq.~(\ref{eq:QnchFit}) may be thought of as accounting
for the quark mass dependence of the pion-cloud source.  This form
then automatically includes the expected heavy quark behaviour where
the $\pi$ and $\eta'$ loop contributions are suppressed.

The form factor, which models the physical structure of the
meson--baryon vertex, characterises the finite size of the pion
source.  Quenched simulations of hadronic charge radii performed at
moderate to heavy quark masses \cite{Leinweber:1991dv} have been
demonstrated to be consistent with experiment once the meson-cloud
properties of full QCD are taken into account
\cite{Hackett-Jones:2000js,Leinweber:1993hj}.  This indicates that the
size of the meson-cloud source is expected to be of similar size in
both quenched and physical QCD. For this reason we proceed to fit both
quenched and physical data with a common value of $\Lambda$.

\begin{figure}[!t]
\begin{center}
\epsfig{file=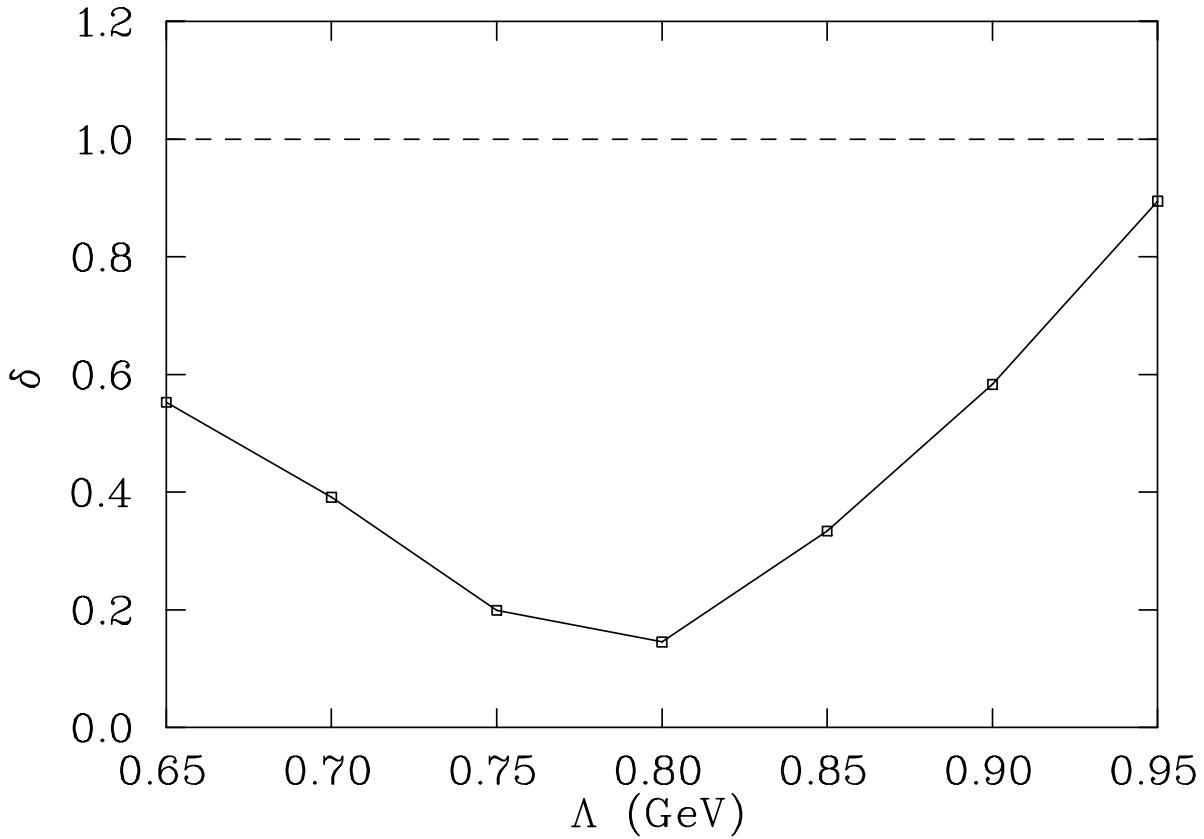, width=11cm, angle=90}
\caption{The value $\delta$ is a measure of the difference
between the quenched and dynamical data sets after accounting for the
relevant self-energy diagrams. This measure is proportional to the
net area contained between the straight lines obtained from the fits
and has been normalised to the case where the self-energy diagrams are
totally neglected. }
\label{fig:varlam}
\end{center}
\end{figure}
For a fixed choice of $\Lambda$, fitting to lattice data amounts to a
linear fit in $\alpha$ and $\beta$.  It turns out that, for a range of
values of $\Lambda$, the values of $\alpha$ and $\beta$ found for the
QQCD data are surprisingly close to the values found for the fit to
dynamical QCD data.  This strongly suggests that the self-energies
included here, which contain the LNA and NLNA behaviour appropriate to
each type of simulation, contain the primary effect of quenching.  To
illustrate the point, Fig.~\ref{fig:varlam} shows a measure, $\delta$,
of the difference between the quenched and dynamical data sets over
the range of $m_\pi$ considered. This measure is proportional to the
net area contained between the straight lines obtained from the fits
and has been normalised to the case where the self-energy diagrams are
totally neglected. The improved agreement between data sets over the
range of dipole masses highlights the effectiveness of this
self-energy correction. It is also worth noting that the $\chi^2/dof$
is also improved by incorporating the self-energies into the fit. For
the preferred dipole mass, $\Lambda=0.8\gev$, this is better by a
factor $2$.

\begin{figure}[!t]
\begin{center}
\epsfig{file=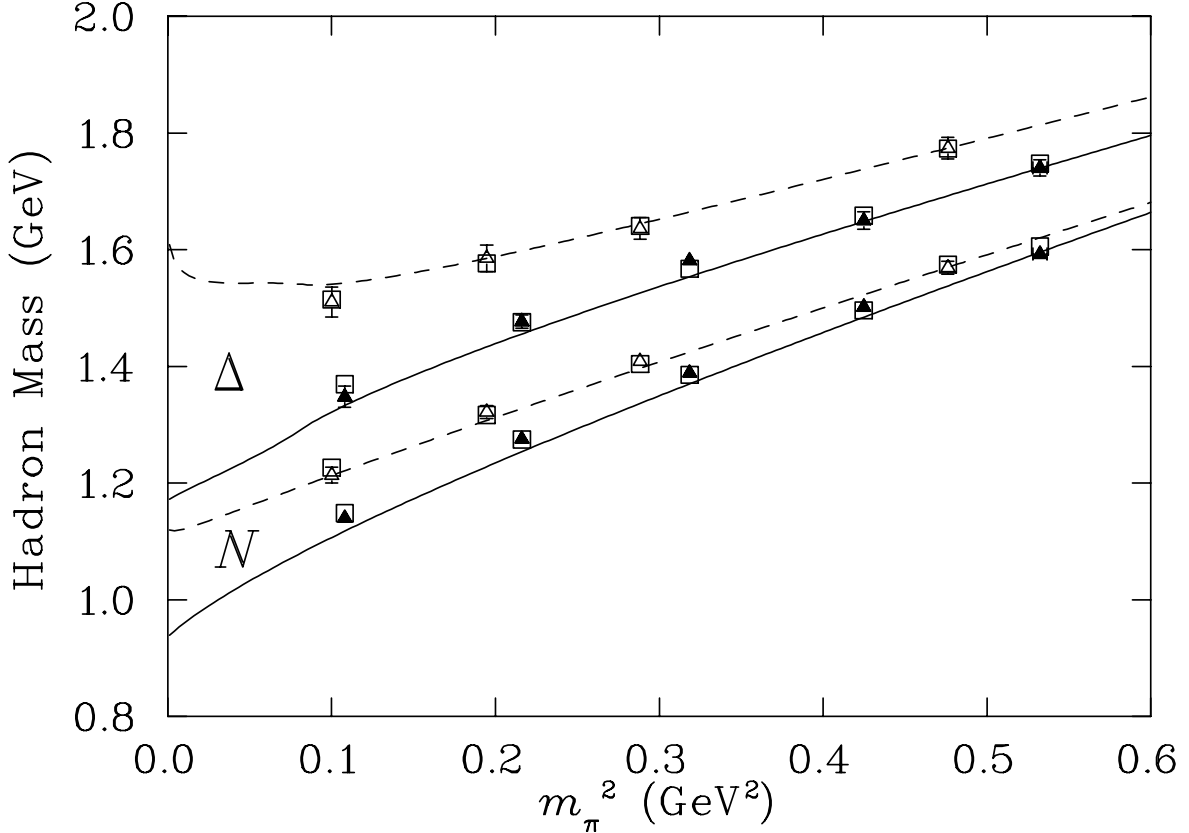, width=11cm, angle=90}
\caption{Fit (open squares) to lattice data
\protect\cite{Bernard:2001av} (Quenched $\vartriangle$, Dynamical
$\blacktriangle$) with adjusted self-energy expressions accounting for
finite volume and lattice spacing artifacts.  The infinite-volume,
continuum limit of quenched (dashed lines) and dynamical (solid lines)
are shown. The lower curves and data points are for the nucleon and
the upper ones for the $\Delta$.}
\label{fig:fqFit}
\end{center}
\end{figure}
Results of both the physical and quenched fits are shown together in
Fig.~\ref{fig:fqFit}. The parameters of the best fits are displayed in
Table~\ref{tab:fitparams}. Here we see the remarkable agreement of the
linear term of our fitting formulae, Eqs.~(\ref{eq:FullFit}) and
(\ref{eq:QnchFit}).  This strongly suggests that the behaviour of the
meson-cloud source is very similar in quenched and full QCD. The
primary difference between the quenched and physical results can then
be described by the meson-loop induced self-energies.

This observation suggests that it may well be possible to make a
connection between quenched simulations and hadron properties in 
the real world. One would fit quenched data with
appropriate self-energies to obtain the linear behaviour of the
meson-cloud source. Then the quenched self-energies would be replaced by
their full-QCD counterparts, hence obtaining more physical results. 
It is clearly very important to test this result further on other
hadrons (e.g. for other members of the octet) and against dynamical
simulations at lower quark masses.
\begin{ruledtabular}
\begin{table}[!t]
\begin{center}
\begin{tabular}{llcccc}
Self-Energy	  & Simulation	& $\alpha_N$  & $\beta_N$       & $\alpha_\Delta$     & $\beta_\Delta$     \\
\hline
Dipole $0.8\gev$  & Physical  & $1.27(2)$     & $0.90(5)$       & $1.45(3)$           & $0.74(8)$          \\
		  & Quenched  & $1.24(2)$     & $0.85(6)$       & $1.45(4)$           & $0.72(11)$         \\
\hline
Nil		  & Physical  & $1.04(2)$     & $1.07(5)$       & $1.28(3)$           & $0.88(8)$          \\
		  & Quenched  & $1.14(2)$     & $0.92(6)$       & $1.44(4)$           & $0.69(11)$         \\
\end{tabular}
\caption{Best fit parameters for both full and quenched data sets with
dipole regulator, $\Lambda=0.8\gev$. The second set correspond to a
simple linear fit, where the self-energy contributions have been
neglected. All masses are in GeV.
\label{tab:fitparams}}
\end{center}
\end{table}

\end{ruledtabular}

\end{section}


\begin{section}{$\Delta$--$N$ Hyperfine Splitting}
The analysis of lattice data has demonstrated the ability to describe
the primary difference between quenched and dynamical simulations in
terms of the meson-loop self-energies. Figure~\ref{fig:hyper} shows
the difference in the self-energy terms for the $N$ and $\Delta$ in
quenched and full QCD, for several values of the common dipole mass.
It is quite clear that there is a difference of between $150$ and
$250\mev$ between the quenched and full QCD cases. Since this
difference was essential in accounting for the clear differences in
the behaviour of the baryon masses in QQCD and full QCD shown in
Fig.~\ref{fig:fqFit}, we have some confidence in using these results
to say how much of the physical $N$-$\Delta$ mass splitting is
associated with pion loops and how much comes from short range
processes, such as gluon exchange.  In fact, an examination of
Fig.~\ref{fig:hyper} for the case of full QCD suggests fairly clearly
that only about $50\mev$ of the observed $300\mev$ $N$-$\Delta$
splitting arises from pion loops. Of course, this result is more
dependent on the assumption of the {\it same} dipole mass parameter at
every vertex than the fits to the $N$ and $\Delta$ masses
individually. Nevertheless, it seems unlikely that more than a third
of the total splitting could come from this source.
\begin{figure}[t]
\begin{center}
\epsfig{file=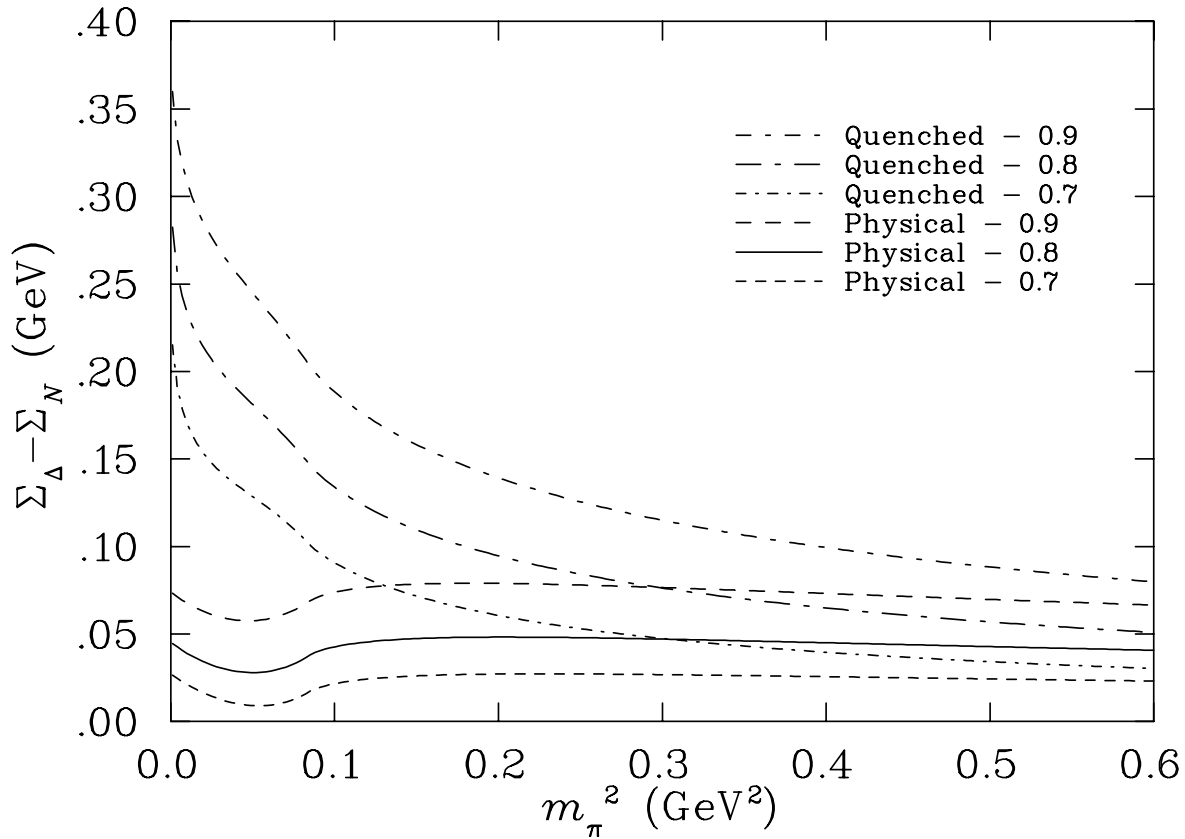, width=11cm, angle=90}
\caption{Meson-loop contribution to the $\Delta$--$N$ mass splitting in
both quenched and full QCD -- for several values of the dipole
mass.}
\label{fig:hyper}
\end{center}
\end{figure}

\end{section}


\begin{section}{Conclusions}
We have investigated the quark mass dependence of the $N$ and $\Delta$
masses within the quenched approximation. The leading chiral behaviour
of hadron masses is known to differ in quenched QCD from the physical
theory.  This knowledge has been used to guide us in the construction
of a functional form which both reproduces the correct chiral
structure, and is consistent with current lattice simulations.  This
procedure of fitting lattice data with a linear term together with the
meson loop corrections which give rise to the LNA and NLNA behaviour
has been demonstrated previously to fit dynamical QCD simulation
results remarkably well. Here we have shown that the application of
the same procedure to quenched results is able to consistently fit the
data in that case as well.  Remarkably, a comparison of the two fits
suggests that the properties of the $N$ and $\Delta$, stripped of
their pion clouds, are essentially the same in quenched and full QCD,
once the scale is set using the Sommer scale appropriate to heavy
quark systems. It is clearly essential to test this finding against
further full QCD simulations at lighter quark masses as well as for
other hadrons.

We have demonstrated that although the quenched approximation gives
rise to more singular behaviour in the chiral limit, this is not
likely to be observed in lattice simulations as these contributions
are quickly suppressed with increasing quark mass.  Indeed our results
suggest that it will be very hard to detect any significant chiral
curvature in the case of the nucleon, while for the $\Delta$ there may
be some small, upward curvature. The $\Delta$--$N$ mass splitting
increases to around $400\mev$ at the physical point in QQCD. As a
consequence of this behaviour, the $\Delta$ mass in the quenched
approximation is expected to differ from the physical mass by
approximately $25\%$. Finally, we have shown that while a
fraction of the physical $N$--$\Delta$ mass difference can be
attributed to a difference in pion self-energy loops, this is unlikely
to amount to more than a third of the observed splitting.

\end{section}


\acknowledgments

We would like to thank S.~Sharpe for many helpful
discussions.  This work was supported by the University of Adelaide
and the Australian Research Council.


\appendix

\begin{section}{Analytic Integration}
Here we summarise analytic expressions for the self-energy
integrals. It should be noted that these expressions are not used in
fitting lattice data. For the purpose of fitting, the continuum
integral is replaced by a discrete sum over the available momenta on
the corresponding lattice, as described in Eq.~(\ref{eq:discrete}).

Firstly we consider the case of the simple meson-loop digram where the
internal baryon line has degenerate mass with the external state,
\begin{equation}
\sigma =  -\frac{3}{16 \pi^2 f_\pi^2} G
	       \int_{0}^{\infty} dk \frac{k^4 u^2(k)}{\omega^2(k)}\, .
\end{equation}
Using a sharp cut-off, $u(k) = \theta (\Lambda - k)$, the integral can
be expressed as
\begin{equation}
\sigma = -\frac{3 G}{16 \pi^2 f_\pi^2} \left( m_\pi^3 
                \arctan\left( \frac{\Lambda}{m_\pi} \right) + 
                \frac{\Lambda^3}{3} - \Lambda m_\pi^2 \right).
\end{equation}
The LNA behaviour of this can then be immediately read from this,
\begin{equation}
\sigma\big|^{\rm LNA} = -\frac{3 G}{32 \pi f_\pi^2} m_\pi^3\, .
\end{equation}
Alternatively, our preferred dipole $u(k) = \Lambda^4/( \Lambda^2 + k^2 )^2$, also provides an analytic
expression for this self-energy,
\begin{equation}
\sigma = -\frac{3 G}{512 \pi f_\pi^2} \frac{\Lambda^5}{(\Lambda + m_\pi)^4} 
              \left( \mpi^2 + 4 \Lambda \mpi + \Lambda^2 \right)\, .
\end{equation}
This gives precisely the same LNA behaviour as the sharp cut-off, as
expected because the non-analytic behaviour is due to the infrared
behaviour of the integral. It is associated with the residue of the
pion propagator pole, and hence independent of an ultraviolet cut-off.

The case of the double meson propagator can also be performed
analytically,
\begin{equation}
\sigma =  \frac{3}{16 \pi^2 f_\pi^2} N
	       \int_{0}^{\infty} dk \frac{k^4 u^2(k)}{\omega^4(k)}\, .
\end{equation}
For $u(k) = \theta (\Lambda - k)$,
\begin{equation}
\sigma = \frac{3}{16 \pi^2 f_\pi^2} N \frac{3 \mpi^2 \Lambda + 2 \Lambda^3 - 3 \mpi (\mpi^2 + \Lambda^2)\arctan\left( \frac{\Lambda}{m_\pi} \right)}{2 (\mpi^2 + \Lambda^2)}.
\end{equation}
For $u(k) = \Lambda^4/( \Lambda^2 + k^2 )^2$,
\begin{equation}
\sigma = \frac{3 N}{512 \pi f_\pi^2} \frac{\Lambda^5 (\mpi + 5 \Lambda)}{(\Lambda + m_\pi)^5} .
\end{equation}
Once again both integrals give the same LNA behaviour,
\begin{equation}
\sigma\big|^{\rm LNA} = -\frac{9 N}{64 \pi f_\pi^2} \mpi\, .
\end{equation}

For the off-diagonal contributions, where the internal baryon is not
degenerate with the external state,
\begin{equation}
\sigma =  - \frac{3}{16\pi^2 \fpi^2} G \int_0^\infty dk \frac{k^4 u^2(k)}
{\omega(k) ( \omega_{BB'} + \omega(k) )},
\end{equation}
with $\omega_{BB'}$ finite. The results do not have a simple form. 
The full expression for the case of a sharp cut-off form
factor can be found in Ref.~\cite{Leinweber:1999ig}. We show the LNA
contribution to this diagram for reference,
\begin{equation}
\sigma\big|^{\rm LNA} = - \frac{9 G}{128\, \pi^2 \fpi^2} 
\frac{\mpi^4}{\omega_{BB'}} \log \mpi\, .
\end{equation}

\end{section}

\begin{section}{Flavour Singlet in Full and Quenched QCD}
\label{app:singlet}
This appendix serves to clarify the derivation of the hairpin
meson-baryon couplings in quenched $\chi$PT.

The flavour singlet $\eta'$ remains light in the quenched
approximation, and is therefore an effective degree of freedom in the
low energy sector. Such excitations must therefore be incorporated
into the low-energy analysis. Within full QCD, resummation of internal
loop diagrams renders the $\eta'$ massive and hence it plays no role
in the low-energy dynamics. For this reason couplings to such flavour
singlet states are neglected. In our analysis, we wish to compare the
low-energy structure of the quenched and physical theories. In this
case, a flavour singlet coupling, like $NN\eta'$, must be included
in the chiral Lagrangian of full QCD in order that it is treated on
equal footing with the quenched theory. This coupling will not alter
any results of the physical theory as any diagram would involve the
propagation of a heavy $\eta'$.

Here we derive best estimates for the flavour singlet couplings in
quenched QCD. This is achieved by comparison of the quenched and full
chiral Lagrangians, under the standard assumption that the couplings
exhibit negligible change between the two theories
\cite{Labrenz:1996jy}. We follow the notation of Labrenz and Sharpe
\cite{Labrenz:1996jy} in the analysis of such contributions. All
symbols retain the same meaning, unless otherwise specified.

The chiral Lagrangian for full QCD can be expressed as
\begin{equation}
\Lg = \Lg_\pi + \Lg_{B\pi} + \Lg_{T\pi}\, .
\end{equation}
The standard octet and decuplet Lagrangians are given with an
additional coupling to an SU(3) flavour singlet state. For clarity we
label the octet and singlet parts of the meson matrix, $A$:
\begin{eqnarray}
\Lg_{B\pi} & = & i \tr ( \bar{B} v\cdot{\cal D} B )           \non    \\
 &  & + 2 D \tr ( \bar{B} S^\mu \{ A^{oct}_\mu , B \} ) 
      + 2 F \tr ( \bar{B} S^\mu [ A^{oct}_\mu , B ] )         \non    \\
 &  & + 2\mu b_D \tr ( \bar{B} \{ {\cal M}^+ , B \} ) 
      + 2\mu b_F \tr ( \bar{B} [ {\cal M}^+ , B ] )           \non    \\
 &  & + 2\mu b_0 \tr ( \bar{B} B ) \tr ( {\cal M}^+ ) 
      + 2 g_s \tr (\bar{B} S^\mu B ) \tr ( A^{sin}_\mu )\, , 
\label{eq:L8b}\\
\Lg_{T\pi} & = & - i \bar{T}^\nu (v\cdot{\cal D}) T_\nu 
     + \Delta M \bar{T}^\nu T_\nu                                          \non \\
 & & + 2 {\cal H} \bar{T}^\nu S^\mu A^{oct}_\mu T_\nu 
     + {\cal C} ( \bar{T}^\nu A^{oct}_\nu B + \bar{B} A^{oct}_\nu T^\nu )  \non \\
 & & + c \bar{T}^\nu {\cal M}^+ T_\nu 
     - \bar{\sigma} \bar{T}^\nu T_\nu \tr ( {\cal M}^+ )                   \non \\
 & & + 2 g_s' \bar{T}^\nu S^\mu T_\nu \tr ( A^{sin}_\mu ) \, .
\end{eqnarray}
The new parameters, $g_s$ and $g_s'$, describe couplings of the
flavour singlet $\eta'$ to baryon octet and decuplet states
respectively. Within full QCD the single vertex has two topologically
different quark flow diagrams as illustrated by the left and
right-hand vertices of Fig.~\ref{fig:etaQF}(a). The left is that of a
$q\bar{q}$ insertion on one of the valence quark lines and the right
is a pure gluonic coupling through a hairpin-style $q\bar{q}$
annihilation. The total coupling is a sum of these two
contributions. Denoting the hairpin vertex coupling by $\gamma_{\rm
QCD}$ and $\gamma_{\rm QCD}'$ for octet and decuplet baryons
respectively we have
\begin{eqnarray}
g_s  & = & \frac{1}{\sqrt{6}}\, g_{\eta'NN}           + \gamma_{\rm QCD}  \, ,  \\
g_s' & = & \frac{1}{\sqrt{6}}\, g_{\eta'\Delta\Delta} + \gamma_{\rm QCD}' \, .
\label{eq:gsin}
\end{eqnarray}
The first of these interactions, $g_{\eta'NN}$
($g_{\eta'\Delta\Delta}$) is related to the axial couplings by SU(6) phenomenology.
We take the standard approach and assign
\begin{eqnarray}
g_{\eta'NN}           & = & \sqrt{2} g_{\eta NN} = \sqrt{\frac{2}{3}} (3 F - D)  \, , \\
g_{\eta'\Delta\Delta} & = & \sqrt{2} g_{\eta\Delta\Delta} = \sqrt{\frac{2}{3}} {\cal H} \, .
\end{eqnarray}

The effective chiral Lagrangian of quenched QCD is \cite{Labrenz:1996jy}
\begin{equation}
\Lg^{(Q)} = \Lg_\Phi^{(Q)} + \Lg_{{\cal B}\Phi}^{(Q)} + \Lg_{{\cal T}\Phi}^{(Q)} \, ,
\end{equation}
where meson and baryon states are now understood to be constructed of
ordinary quarks and bosonic quarks.  The general Lagrangian for the
heavy fields can be written in terms of the rank-three tensor fields
as defined in Ref.~\cite{Labrenz:1996jy}, $\cal B$ and $\cal T$,
\begin{eqnarray}
\Lg_{{\cal B}\Phi}^{(Q)} & = & i ( \bar{{\cal B}} v\cdot{\cal D} {\cal B} )    \non    \\
 &  & + 2 \alpha ( \bar{{\cal B}} S^\mu {\cal B} A_\mu ) 
      + 2 \beta ( \bar{B} S^\mu A_\mu {\cal B} )                               \non    \\
 &  & + 2 \gamma_s ( \bar{{\cal B}} S^\mu {\cal B} ) \str ( A_\mu ) 
      + \alpha_M ( \bar{{\cal B}} {\cal B} {\cal M}^+ )                        \non    \\
 &  & + \beta_M ( \bar{{\cal B}} {\cal M}^+ {\cal B} ) 
      + \sigma ( \bar{{\cal B}} {\cal B} ) \str ( {\cal M}^+ ) \, ,
\label{eq:LQ8b}\\
\Lg_{{\cal T}\Phi}^{(Q)} & = & 
     - i ( \bar{{\cal T}}^\nu (v\cdot{\cal D}) {\cal T}_\nu ) 
     + \Delta M ( \bar{{\cal T}}^\nu {\cal T}_\nu )                           \non    \\
 & & + 2 {\cal H} ( \bar{{\cal T}}^\nu S^\mu A_\mu {\cal T}_\nu ) 
     - \sqrt{\frac{3}{2}} {\cal C} [ \bar{{\cal T}}^\nu A_\nu B 
     + \bar{B} A_\nu {\cal T}^\nu ]                                           \non    \\
 & & + 2 \gamma_s' ( \bar{{\cal T}}^\nu S^\mu {\cal T}_\nu ) \str ( A_\mu )   \non    \\
 & & + c \bar{{\cal T}}^\nu {\cal M}^+ {\cal T}_\nu 
     - \bar{\sigma} ( \bar{{\cal T}}^\nu {\cal T}_\nu ) \tr ( {\cal M}^+ ) \, .
\label{eq:LQ10b}
\end{eqnarray}
It should be noted that the terms $\gamma_s$ and $\gamma_s'$ describe
both types of flavour-singlet coupling, not just that arising through
the hairpin alone. Similarly to Eq.~(\ref{eq:gsin}), in the quenched
theory these can be described by
\begin{eqnarray}
\gamma_s  & = & \frac{1}{\sqrt{6}}\, g_{\eta'NN}           + \gamma  \, ,  \\
\gamma_s' & = & \frac{1}{\sqrt{6}}\, g_{\eta'\Delta\Delta} + \gamma' \, ,
\end{eqnarray}
where the terms $\gamma$ and $\gamma'$ now correspond to the pure
hairpin couplings as used in Ref.~\cite{Labrenz:1996jy}. Here we also
note the the terms $g_{\eta'NN}$ and $g_{\eta'\Delta\Delta}$ are
unchanged in going to the quenched theory, this is consistent with the
assumption that the chiral parameters $F$ and $D$ are unchanged
between the two theories.

One can then relate the quenched chiral Lagrangian back to that of
full QCD by restricting the indices on the tensor fields, $\cal B$ and
$\cal T$, to those corresponding to the physical quarks. The details
of this procedure are described in Ref.~\cite{Labrenz:1996jy}.
Performing these restrictions on the octet-baryon, quenched chiral
Lagrangian (Eq.~\ref{eq:LQ8b}) one finds:
\begin{eqnarray}
\Lg_{{\cal B}\Phi}^{(Q)} |_R & = & i ( \bar{{\cal B}} v\cdot{\cal D} {\cal B} ) |_R                                      \non \\
 & & + \frac{2}{3} ( 2 \alpha - \beta ) \tr ( \bar{B} S^\mu A_\mu B )
     + \frac{1}{3} ( - \alpha - 4 \beta ) \tr ( \bar{B} S^\mu B A_\mu )                      \non \\
 & & + \frac{1}{3} ( \alpha + 4 \beta + 6 \gamma_s ) \tr ( \bar{B} S^\mu B ) \tr ( A_\mu )   \non \\
 & & + \frac{1}{3} ( 2 \alpha_M - \beta_M ) \tr ( \bar{B} {\cal M}^+ B )
     + \frac{1}{6} ( - \alpha_M - 4 \beta_M ) \tr ( \bar{B} B {\cal M}^+ )                   \non \\
 & & + \frac{1}{6} ( \alpha_M + 4 \beta_M + 6 \sigma ) \tr ( \bar{B} B ) \tr ( {\cal M}^+ ) \, .
\end{eqnarray}
Equating this with Eq.~(\ref{eq:L8b}) gives:
\begin{eqnarray}
\frac{2}{3} ( 2 \alpha - \beta )                & = & 2 D + 2 F \label{eq:ab1} \, , \\
\frac{1}{3} ( - \alpha - 4 \beta )              & = & 2 D - 2 F \label{eq:ab2} \, , \\
\frac{1}{3} ( \alpha + 4 \beta + 6 \gamma_s )   & = & 2 g_s   \label{eq:gdef}  \, , \\
\frac{1}{3} ( 2 \alpha_M - \beta_M )            & = & 2\mu b_D + 2\mu b_F      \, , \\
\frac{1}{6} ( - \alpha_M - 4 \beta_M )          & = & 2\mu b_D - 2\mu b_F      \, , \\
\frac{1}{6} ( \alpha_M + 4 \beta_M + 6 \sigma ) & = & 2\mu b_0                 \, .
\end{eqnarray}
In extracting the flavour-singlet part, Eq.~(\ref{eq:gdef}) provides us with
\begin{equation}
\frac{1}{3} \alpha + \frac{4}{3} \beta + \sqrt{\frac{2}{3}}\, g_{\eta'NN} + 2 \gamma = \sqrt{\frac{2}{3}}\, g_{\eta'NN} + 2 \gamma_{\rm QCD} \, ,
\end{equation}
and combining with Eqs.~(\ref{eq:ab1},\ref{eq:ab2}) one arrives at
\begin{equation}
\gamma = \gamma_{\rm QCD} + D - F \, .
\end{equation}
The restrictions are much simpler for the decuplet case and one finds
\begin{equation}
\gamma' = \gamma_{\rm QCD}' \, .
\end{equation}

In estimating the hairpin-type couplings in full QCD one assumes that
they are relatively small, $\gamma_{\rm QCD}\ll g_{\eta' NN}$, due to
OZI-type suppression \cite{Close:1979bt}. With analogous arguments for
the decuplet, we take $\gamma_{\rm QCD}=\gamma_{\rm QCD}'=0$.  We do
note that the U(1) axial anomaly may be effective in overcoming the
OZI rule in the case of $\eta'$ couplings \cite{Bass:1999is}, but as
we mentioned in the text the main conclusions of our present analysis
are not very sensitive to the precise value of the $\eta'$-nucleon
coupling.

\end{section}


\bibliography{extrap,chpt,general,lattice,chiral}
\bibliographystyle{apsrev}

\end{document}